\documentclass[aps,prd,twocolumn,preprintnumbers,amsmath,amssymb,groupedaddress]{revtex4}

\usepackage[dvips]{graphicx}
\usepackage{dcolumn}
\usepackage{bm}

\def\Journal#1#2#3#4{{\em #1} {\bf #2}, #3 (#4).}

\begin{document}



\title{Challenges in signal analysis of resonant-mass gravitational wave detectors }

\author{Nadja Sim\~ao Magalh\~aes}


\affiliation{Centro Federal de Educa\c{c}\~ao Tecn\'{o}logica de
S\~ao Paulo, Rua Pedro Vicente 625, S\~ao Paulo, SP 01109-010,
Brazil and \\
Instituto Tecnol\'ogico de Aeron\'autica - Departamento de
F\'{\i}sica\\ P\c ca. Mal. Eduardo Gomes 50, S\~ao Jos\'e dos
Campos, SP 12228-900, Brazil}


\begin{abstract}
An overview of the main points related to data analysis in
resonant-mass gravitational wave detectors will be presented.
Recent developments on the data analysis system for the Brazilian
detector SCHENBERG will be emphasized.
\end{abstract}


\maketitle

\section{Introduction}

As predicted by the theory of relativity and other theories of
gravitation, time-dependent gravitational forces are expected to
propagate in spacetime in the form of waves \cite{mis73} . Such
gravitational radiation is extremely difficult to detect because
gravitation is the weakest of all the fundamental forces of
nature.

For instance, a wave of very strong amplitude could generate a
displacement of $10^{-18}$m in a system (``antenna") with typical
length of $1$m. In order to detect such a tiny displacement
special sensors must be used and the signal should be sent to
computers to be properly analyzed. The path between the antenna
and the computer is tricky, though, because many spurious signals
- noise - come as well.

\begin{figure}
    \begin{center}
        \includegraphics[width=3.5cm]{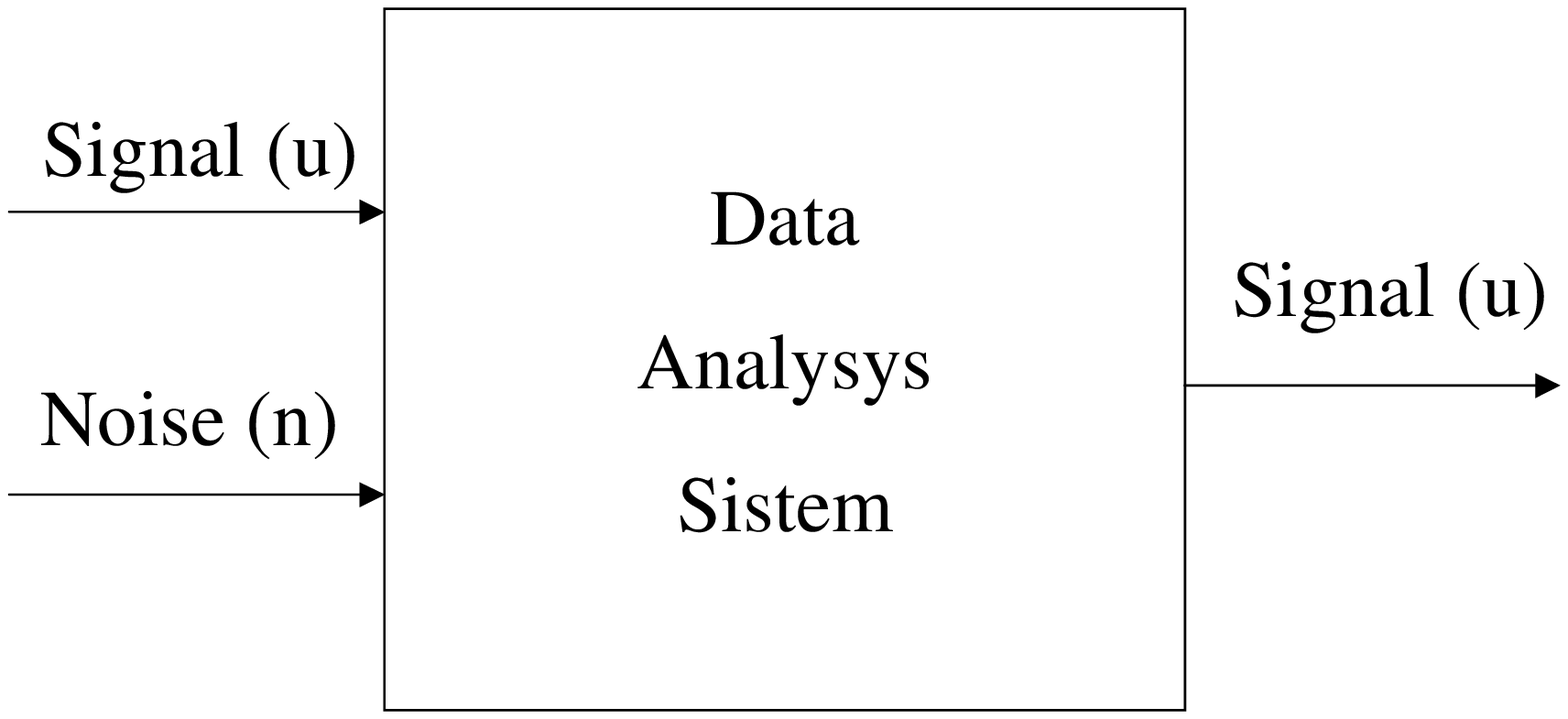}
        \caption{The main objective of signal analysis is to
retrieve the useful signal out of noise.} \label{fig:dan}
    \end{center}
\end{figure}

When an experiment is performed, normally the signal that is the
object of the observation (the ``useful" signal, $u(t)$) is
accompanied by other, unwanted signals labelled with the generic
name of ``noise" ($n(t)$). The goal of signal analysis is to
retrieve the useful signal out of noise, as illustrated in Figure
\ref{fig:dan}. In the case of gravitational wave experiments, the
useful signal is a gravitational wave.

Because signal analysis identifies the signal in the midst of the
noise it becomes a fundamental part of the experiment. To do so it
is necessary to know as much as possible about the signal and the
sources of noise. In this work general features of data analysis
in gravitational wave resonant-mass detectors will be presented,
so that the reader can become familiar with the latest challenges
faced by the Brazilian group that works in the SCHENBERG
gravitational wave detector within this important field.

This paper is organized as follows: in Section \ref{sec:sources}
the motivation for the research is presented, namely gravitational
waves and their sources. Section \ref{sec:det} is devoted to
presenting general features of resonant-mass detectors, while
Section \ref{sec:noi} discusses some noises present in such
detectors. Several basic concepts usual in the context of data
analysis are presented in Section \ref{sec:basicanalysis} while
some important challenges that must be faced in the analysis of
SCHENBERG's data are discussed in Section \ref{sec:datasch}. The
closing comments are made in the final section.

\section{Gravitational wave sources} \label{sec:sources}

Strong gravitational waves (g.w.) are expected to be generated by
astrophysical objects \cite{thornebook}. For instance, two very
massive stars orbiting each other would emit such waves. In
particular, when they are coalescing they emit waves in a large
frequency range. Another example is given by a black hole that
rings down, also emitting in different wavelengths \cite{ces04}.

Astrophysical sources emit basically three kinds of waves,
depending on the waveform: bursts (impulsive signals of short
duration, like those produced by supernova explosions), continuous
waves (periodic, with long duration, like those emitted by stable
binary systems) and stochastic waves (a spectrum composed with the
superposition of many sources, as the one expected from
cosmological origin).

From the analysis of the detected waves emitted by such sources
important information is expected. The very first direct detection
will provide a test for one of the predictions of the theory of
general relativity. Then, continuous observation will allow
testing other theories of gravitation \cite{maga96,merk98},
besides initiating gravitational astronomy \cite{maga95,schu05}.

In order to make astrophysical observations the following
parameters that characterize the g.w. are needed: the amplitudes
of the two states of polarization of the wave as functions of time
($h_+ (t)$ and $h_{\times}(t)$), the source direction in the sky
(given by the angles $\theta$ and $\phi$) and the phase of the
wave (obtainable from the detailed time dependency of $h_+ (t)$
and $h_{\times}(t)$, usually associated to the polarization angle
$\alpha$). Therefore, gravitational wave observatories must be
able to detect at least five independent observables in order to
allow for gravitational wave astronomy to start. On the other
hand, detection of at least one observable would be a strong
evidence of the existence of such waves. All the existing
detectors are presently aimed at this first detection, while being
prepared to become part of g.w. observatories in the future as
well. In the next section a brief overview on resonant-mass
detectors will be presented.

\section{Resonant-mass detectors} \label{sec:det}

\begin{figure}
    \begin{center}
        \includegraphics[width=4cm]{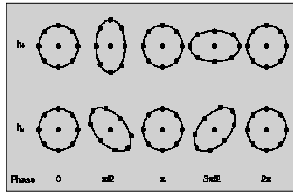}
        \caption{When a gravitational wave passes through a
ring of particle it changes their relative positions, depending on
the wave's polarization. The top line shows the motion produced by
a wave with polarization ``+", while the bottom line shows the
motion due to a wave with polarization ``$\times$".}
\label{fig:ring}
    \end{center}
\end{figure}

When a gravitational wave passes through a ring of particles it
changes their relative positions \cite{schutzbook}, as shown in
Figure \ref{fig:ring}. Similarly, solid bodies are distorted in
the presence of such waves due to the changes in spacetime. This
effect is the basis for resonant-mass gravitational wave
detectors. For instance, a massive cylinder would oscillate
longitudinally in the presence of a g.w., in a frequency resonant
with the wave.

The first of such detectors was built in the 1960's by Joseph
Weber. It was a massive, cylindrical aluminum bar at room
temperature, $1.5$m long, monitored by piezoelectric transducers.
Nowadays there are improved bar detectors, sometimes cooled down
to millikelvin temperatures, longer and using more sensitive
transducers \cite{amaldi6}. Those that are operational at the
moment are: ALLEGRO (EUA), EXPLORER (Switzerland), NAUTILUS and
AURIGA (both in Italy)

Bar detectors are able to determine only one observable, and due
to their geometry there are directions in which they are more
sensitive than others. For these reasons several bar detectors,
appropriately positioned, would have to be used if one wished to
build a gravitational observatory with only this kind of detector.
In fact, investigations have already been done in this direction
\cite{igec}.

Besides cylindrical geometry, it has been known for some time that
spherical, solid objects could be used as g.w. resonant-mass
detectors. In principle this geometry has no preferred direction
of observation (i.e., it is omnidirectional) and the five
observables needed for gravitational astronomy could be obtained
from only one detector appropriately equipped
\cite{maga95,maga97b}. These are major advantages, but for many
years bars were preferred because they were easier to machine and
equip.

\begin{figure}
    \begin{center}
        \includegraphics[width=5cm]{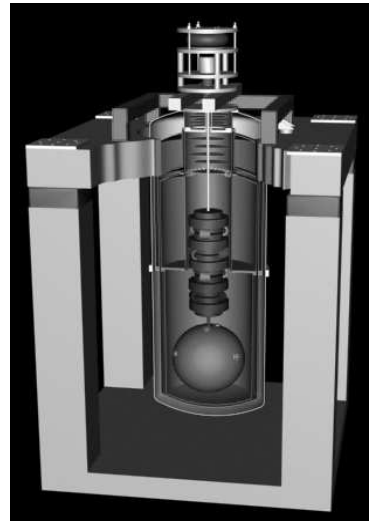}
        \caption{Schematics of the SCHENBERG gravitational wave detector.}
\label{fig:schen}
    \end{center}
\end{figure}

Under the rationale that spherical geometry is the best, in the
last decade a lot of effort has been invested in investigating and
building resonant-mass g.w. detectors with this design
\cite{omni1,amaldi6}. The Brazilian {\it Mario SCHENBERG} detector
is one of these last generation detectors \cite{ody100ys} (see
Figure \ref{fig:schen}). When fully operational it will be able
not only to acknowledge the presence of a g.w. within its
bandwidth: it will be able to inform the direction of its source
in sky, the wave's amplitude and its polarization - one only
antenna working as a gravitational wave observatory in a bandwidth
between $3000$ and $3400$Hz, sensitive to displacements around
$10^{-20}$m . To this end at least 6 transducers will continuously
monitor displacements of the antenna's surface. The data collected
will be sent to be analyzed and, as mentioned above, it will
certainly be accompanied by some noise. In the next section some
of the main noise sources in SCHENBERG will be presented.

\section{Noise sources in SCHENBERG} \label{sec:noi}

People walking around the detector may generate undesired
vibrations on the resonant-mass antenna which become {\it seismic
noise} in the signal detected by the transducers. This kind of
noise can be minimized with the use of an appropriate suspension
system, as the one used in SCHENBERG \cite{ody02, ody100ys}: the
spherical antenna is carefully suspended from its center of mass
by a rod attached to a system of ``springs" specially designed to
attenuate seismic noise.

Other kinds of noise are harder to minimize. The Brownian motion
of the atoms of the antenna generate {\it thermal noise}, which
can be reduced by cooling the massive sphere to temperatures as
low as possible. SCHENBERG is expected to be cooled down to $4.2$K
(liquid Helium temperature) in its first test run. After the
antenna is cooled down, the standard procedure to reduce its
thermal noise has been to apply a threshold in the detected signal
above which a signal may be considered a candidate for a g.w.
event. It has been typical to set the threshold to amplitude
signal-to-noise ratio between 3 and 5 \cite{igec03}. In the future
it is planned to cool SCHENBERG down to temperatures of the order
of millikelvin, decreasing the antenna's thermal noise in about
one order of magnitude \cite{fra04}.

Other relevant noises in SCHENBERG are related to the transducers
used to translate the mechanical motion of the antenna into an
electrical signal. They are microwave parametric transducers
\cite{andrade04} that generate noises that can be divided into two
groups: narrow band and broadband noises. The narrow band noise
includes the back action and the Brownian noises. The broadband
noise can be divided into two components: one due to the amplifier
and another due to the phase noise in the pump microwave source.
All these noises have been modelled mathematically \cite{fra04} so
that the corresponding expressions can be used in the development
of digital filters that help to minimize these noises in the
detected signal. The importance of the determination of such
mathematical expressions will become more clear within the context
of data analysis, as follows.

\section{Basic concepts in data analysis} \label{sec:basicanalysis}

The electric signal that leaves the transducers carries
information of both the gravitational wave and the different
noises. Data analysis then tries to extract the g.w. signal as
clean as possible out of noise. In other words, one is concerned
with improving the signal-to-noise ratio of measurements: an
accurate measurement can be made when the g.w. signal causes an
output of the detector that is large compared to the random
variations of the output when no g.w. signal is present. In what
follows some basic concepts important do data analysis will be
presented, including signal-to-noise ratio. The theory of signal
detection, much of it invented in the context of the development
of the radar during World War II, can be found in a number of
books \cite{whalenbook,zubakovbook}.

The output of a g.w. detector is expected to be continuously
recorded, and this allows one to know this output as a function of
time: $s(t)$. Such mathematical object is called a {\it time
series}. In principle one hopes to know as much of the g.w. signal
as possible, making it a deterministic time series of a
predetermined form, $u(t)$, or a {\it template}. For instance,
burst signal are commonly associated to Dirac's delta function. On
the other hand, noise is usually associated to time series,
$n(t)$, that randomly varies from one realization to the next.
Before data analysis one then has $s(t) = u(t)+n(t)$. Notice that
in the absence of any g.w. signal the detector output is just
plain noise: $s(t) = n(t)$.

These distinctions between the time series of the g.w. signal and
noise imply different mathematical operations to characterize the
regularities of these series. A useful way to express the time
series is in terms of its Fourier transform, $S(f)$, a function of
frequency which contains the same information of the time series
$s(t)$, defined by
\[
S(f) \equiv \frac{1}{\sqrt{2 \pi}}\int^{\infty}_{-\infty} s(t)
e^{-2\pi f t} dt.
\]
This puts the data in the {\it frequency domain}. Notice that
since data is sampled in discrete sets in the practical world, the
discrete analog of the above equation is a more appropriate way to
define the Fourier transform for real data sets. In fact, there
exists a powerful algorithm for calculating discrete Fourier
transforms, the FFT or Fast Fourier Transform, which is very
useful in modern laboratory instrumentation. But for this brief
presentation of the subject the continuous approach will be
preferred.

The value of the Fourier transform of $s(t)$ at the frequency $f$
is a measure of the degree to which $s(t)$ varies like a sinusoid
of frequency $f$. Putting it in a better way, the Fourier
transform gives the contribution of a sinusoid of frequency $f$ to
a sum of sinusoids that equals the function of interest, $s(t)$ in
this example.

For a deterministic signal as the one expected for g.w. its
Fourier transform can be calculate quite straightforwardly from
the above definition. The characterization of a random time series
(noise) in the frequency domain demands an extra step, which is
the definition of the {\it autocorrelation function} of the noise:
\[
n * n(\tau )\equiv \int^{\infty}_{-\infty} n(t) n(t+\tau ) dt.
\]
This function of a time offset $\tau $ is a way of measuring how
$n(t)$ is related to itself, at different time offsets between two
copies of $n(t)$. When $\tau = 0$ the time series is aligned to
itself so the autocorrelation function will always have a maximum
at this time offset. The width of $n \star n$ indicates how
rapidly the noise changes with time.

The translation of the random time series $n(t)$ into the
frequency domain is then given by the the {\it power spectrum} of
this series (also known as power spectral density), defined as the
Fourier transform of the autocorrelation function:
\[
P_s(f)\equiv \frac{1}{\sqrt{2\pi }}\int^{\infty}_{-\infty} n *
n(\tau) e^{-2\pi f \tau} d\tau.
\]
$P_s(f)$ is a measure of the amount of time variation in the time
series that occurs in the specific frequency $f$.

Instead of thinking in terms of complex integrals of both positive
and negative frequencies, experimentalists often prefer to think
in terms of sines and cosines of positive frequencies. So it is
common to use the {\it single sided power spectrum}, $s^2(f)$,
defined by
\[
s^2 \left( f \right) \equiv \left\{ {\begin{array}{*{20}c}
   {2P_s \left( f \right),{\kern 1pt} \;if\;f \ge 0}  \\
   {0,\quad \quad otherwise.}  \\
\end{array}} \right.
\]
Often the expression ``power spectrum" refers to this definition
within the experimental context. Also, an object derived from this
quantity is commonly used, the {\it amplitude spectral density},
defined simply by $s(f)\equiv \sqrt{s^2(f)}$. Details on the
advantages and disadvantages of the use of these quantities can be
found in \cite{saulsonbook}.

After presenting how signal and noise can be characterized in the
theory of signal detection, now the concept of signal-to-noise
ratio (or SNR) will be introduced. This is a dimensionless figure
of merit for a measurement. In order to create such a
dimensionless quantity one must keep in mind that signal detection
is the process of searching for a pattern resembling the template
in the middle of a noisy record. Such a pattern should occur with
a strength unlikely to be due to noise alone.

The match between the time record and the template can be
estimated by the {\it cross-correlation integral} between the
template $u(t)$ and the time record $s(t)$, evaluate for all
possible times at which the signal could have arrived:
\[
s * u(t) \equiv \int^{\infty}_{-\infty} s(\tau) u(t+\tau) d\tau.
\]
This definition is similar to the definition of the
autocorrelation function, given previously. The cross-correlation
function indicates how related the functions $s(t)$ and $u(t)$ are
 to each other. Then one way to characterize the strength
$S^2$ of the signal present in any time $t$ is using the
cross-correlation between the expected form of the output (the
template $u(t)$) and the output $s(t)$:
\[
S^2\equiv |s*u(t)|.
\]

As for the noise, it can be characterized by $N^2$, the mean
square value of the cross-correlation between the output $s(t)$ in
the absence of g.w. (i.e., noise) and a given template:
\[
 N^2  \equiv  {\left\langle {\left( {s * u\left( t \right)}
\right)^2 } \right\rangle }.
\]
The brackets $\left\langle \right\rangle$ indicate averaging over
time.

With these characterizations one defines the signal-to-noise ratio
as the square root of the ratio of the measure of the amount of
signal present ($S^2$) to the expected value due to noise alone
($N^2)$:
\[ SNR \equiv \sqrt {{{S^2 } \mathord{\left/
 {\vphantom {{S^2 } {N^2 }}} \right.
 \kern-\nulldelimiterspace} {N^2 }}}.
\]
A large SNR indicates that something is present in the time series
$s(t)$ other than noise. In practice $SNR\approx 1$ is not of much
use, but $SNR \gtrsim 10$ indicate detection of some confidence.
Therefore the goal is to maximize SNR in order to detect a g.w.
This can be accomplished in a number of ways. For instance, $n(t)$
for antenna's thermal noise can be reduced by cooling the antenna
down. As for transducer's noise, a digital filter is useful.
Filtering is such an important part of signal analysis that it
will be briefly reviewed next.

Suppose there is a device possessing one single input ($i(t)$) and
a single output($s(t)$). Such device will be considered a {\it
linear system} if there is some linear relationship between the
input and the output: $s(t) = a \, i(t)$. When the relationship
between the input and the output does not change with time, then
the device is a {\it linear time-invariant system} (or just linear
system for short).

The {\it filters} considered here are linear time-invariant
devices in which the input and output are quantities with the same
dimensions. On the other hand, the term {\it transducer} is used
as a general name for a linear system whose input and output have
different physical units.

One way to specify the input-output relationship in a linear
system is to give the ``impulse response", $g(t)$. This function
is the output obtained when a single unit impulse is applied to
the input at $t=0$. In the frequency domain the Fourier transform
of the impulse response, $G(f)$, is called the {\it frequency
response} (or sometimes {\it transfer function}). This is a
complex-valued function of the frequency $f$ whose real part
represents the response ``in phase" with a sinusoidal input of
frequency $f$, while the imaginary part corresponds to the
``quadrature" component. One can show that if $I(f)$ is the
Fourier transform of the input and $S(f)$ is the Fourier transform
of the output, then $G(f)=S(f)/I(f)$. This equation implies also
that in Fourier space the output of a linear system is simply the
product of the input and the frequency response, with no need to
calculate convolution integrals.

\section{Aspects of SCHENBERG's Data Analysis} \label{sec:datasch}

The theory presented in the last section shows that the knowledge
of SNR for SCHENBERG demands the determination of the time series
of: the detector's output, the g.w. signal (template) and the
noise.

In the particular case of SCHENBERG it is necessary to combine the
outputs of several transducers to create the time series of the
output, $s(t)$. This combination is part of the mathematical
modelling of the detector. There are 6 transducers planned to
monitor the antenna surface's motion and there are mathematical
models for the detector for the case that all these transducers
operational. Such models have investigated two situations: one in
which the transducers are perfectly uncoupled
\cite{maga97b,merk97,cesartese} and another in which the
transducers are somehow coupled to each other \cite{merk99}, an
instance that still offers possibilities of investigation.

One of the challenges presently faced by the data analysis group
within the GRAVITON project \cite{gravitonhome} (the one SCHENBERG
is part of) is to develop a model of the detector with less than 6
transducers. In this case there is a break in the convenient
buckyball symmetry \cite{merk97} and the consequences of this fact
must be investigated. Two approaches are now under investigation:
one considers the model already developed for 6 transducers
\cite{cesardissert} and simply reduces their number; the other
considers the fewer transducers as independent devices and
redesign the mathematical model. For the study of the last
approach several references in the literature may be used as a
starting point
\cite{pizzela88,gurseltinto89,cerdonio93,zhoumichelson95,pai02}.

An actual time series of the detector's output is expected be
known as soon as it is collecting data, in the next months. At
least three transducers are expected to be installed then, so the
investigation mentioned above should be conclude soon.

The work on the knowledge of the time series of the template has
started long ago and is one of the more active fields in data
analysis \cite{gwdaw04}, so that several templates already exist
that can be used \cite{ces04,ody04,araujo04}. But still many
challenges remain to be faced in the field of g.w. sources. At the
moment one of the investigations been carried out by SCHENBERG's
data analysis group in this way refers to the detection of
astrophysically unmodelled bursts of gravitational radiation using
wavelets \cite{mag_enfpc05}.

A significant work has already been done to determine the time
series of the noise in SCHENBERG, as mentioned in Section
\ref{sec:noi}. Since all time series needed are available one is
able to translate them into the frequency domain and determine
SCHENBERG's SNR, as is done in \cite{ces04}. Also, based on the
model with 6 transducers operational simulations of the detector
in the presence of noise have already been made (see Costa and
Aguiar in \cite{amaldi6}).

The improvement of SNR can be accomplished by using appropriate
digital filters, for instance. When noise is white (broadband,
spread over spectrum and stationary, implying a spectral density
that does not depend on the frequency, which is commonly the case)
it can be shown that the best filter for a given template is the
{\it matched filter}. This linear system has an impulse response
which is the time reverse of the signal one is interested in:
$g(t)= s(-t)$. This filter is presently under investigation for
the case of SCHENBERG within GRAVITON's data analysis group.

When noise is not white (either by not being broadband nor
stationary, or both) other strategies are used. For instance, the
monitoring of the environment with seismographers help vetoing
seismic noises. Monitors for cosmic rays work similarly. The
possible influence of lightings on the data has been investigated
as well (see Magalhaes, Marinho, Jr. and Aguiar in
\cite{amaldi6}).

In the particular case of SCHENBERG, which will be monitored by
several transducers simultaneously, one may wonder if filtering
should be performed at the transducers outputs to veto non-white,
non-stationary noise before combining them to extract the wave's
parameters. This is the case when several bar detectors work in
coincidence \cite{igec03}, spaced around the world with some
different characteristics among themselves. Maybe such a procedure
would eliminate some noise due to local disturbances in individual
transducers so that the combined data could be less noisy.

However it can be argued \cite{ody_private} that since SCHENBERG's
identical transducers are collecting data simultaneously at the
same site there is no need to risk loss of information by
filtering their data before combining them, a combination that
will be possibly done in real time and with all characteristics
under control. The data analysis system is then expected to be
able to optimize SNR using the combined outputs without
intermediate filtering.

Besides the matched filter another kind of filter is under
investigation for SCHENBERG, namely an {\it adaptative filter},
one that changes with the variations of the power spectrum of the
noise. This is a device useful in the presence of non-stationary
noise, like electric and seismic noises. This kind of filter has
already been investigated within the g.w. detection context
\cite{broccofrasca04}.

Finally, an aspect that still deserves investigation is the use of
the Bayesian statistics in SCHENBERG's data analysis. Such kind of
study is already been carried out related to other detectors (see
L.S. Finn in \cite{amaldi6}).

\section{Concluding Remarks}

In this work a brief overview of data analysis in resonant-mass
gravitational wave detectors was presented, with emphasis on
issues involving the SCHENBERG detector. This detector, installed
at the Physics Institute of the University of Sao Paulo (Sao Paulo
city), is expected to be collecting data soon. It will be able to
run in coincidence with other detectors around the world,
particularly the broadband interferometric ones (see links for the
several groups at \cite{gravitonhome}). Such kind of coincidence
is important in the first place to increase the credibility on the
detection of a g.w. For this reason it is interesting to consider
the development of a common protocol for information exchange
between SCHENBERG and these detectors.

It is worth pointing out the particular feature of SCHENBERG of
working as an observatory of g.w. by itself due to its capability
of several simultaneous measurements. Monitored by three
transducers, as it is planned for the near future, this detector
will be able to determine the squared amplitude and the direction
of propagation of a g.w. sufficiently strong within its frequency
band. Only spherical detectors like SCHENBERG are able to have
such versatility. This is the case of the MiniGrail, another
spherical detector built in The Netherlands (see de Waard in
\cite{amaldi6}) sensitive to frequencies smaller than SCHENBERG's.
Also, an Italian group is interested in building a large spherical
detector (see V. Fafone in \cite{amaldi6}).

There is the belief within the international community that works
with g.w. detection that the first direct detection of gravitation
radiation from an astrophysical source will become a reality in
the near future. This will open a new window to the universe,
bringing new information about known objects and about fairly
unknown things, like dark matter. In order to extract such
information from the huge amount of data that is expected to be
generated from the detectors' outputs (which has already started)
a lot of work will be demanded in the field of data analysis. This
is a promising field.

\section*{Acknowledgements}
The author is thankful to Odylio D. Aguiar for kindly allowing the
reproduction of Figure \ref{fig:schen}.






\end{document}